\documentclass[aps,prx,10pt,amsmath,amssymb,floatfix,superscriptaddress,notitlepage,twocolumn]{revtex4-2}
\usepackage[T1]{fontenc}
\usepackage{hyperref}
\usepackage{graphicx}
\usepackage{dcolumn}
\usepackage{bm}
\usepackage{xcolor}
\usepackage{braket}
\usepackage{amsfonts}
\usepackage{amssymb}
\usepackage[normalem]{ulem}
\usepackage{dsfont}
\usepackage{lineno}
\usepackage{soul}
\usepackage{bbold}
\usepackage{booktabs}
\usepackage{siunitx}

\newcommand\qfreq{|\boldsymbol{\omega}_L|/2\pi}
\newcommand\gperp{\boldsymbol{\lambda}_\perp}
\newcommand\normgperp{|\boldsymbol{\lambda}_\perp|/2\pi}

\newcommand{\gt}{\hat{g}}

\begin{document}

\title{Switchable spin-photon coupling with hole spins in single-quantum dots}

\author{Carlos Sagaseta}
\affiliation{Departamento de Física, Universidad Carlos III de Madrid, Avda. de la Universidad 30, 28911 Leganés, Spain}
\affiliation{Instituto de Ciencia de Materiales de Madrid (ICMM), Consejo Superior de Investigaciones Científicas (CSIC), Sor Juana Inés de la Cruz 3, 28049 Madrid, Spain}
\author{M. J. Calder\'on}
\affiliation{Instituto de Ciencia de Materiales de Madrid (ICMM), Consejo Superior de Investigaciones Científicas (CSIC), Sor Juana Inés de la Cruz 3, 28049 Madrid, Spain}
\author{Jos\'e C. Abadillo-Uriel}
\email{jc.abadillo.uriel@csic.es}
\affiliation{Instituto de Ciencia de Materiales de Madrid (ICMM), Consejo Superior de Investigaciones Científicas (CSIC), Sor Juana Inés de la Cruz 3, 28049 Madrid, Spain}%

\newcommand{\jc}[1]{{\color{red} #1}}

\date{\today}

\begin{abstract}
Spin qubits in semiconductor quantum dots offer a gate-tunable platform for quantum information processing. While two-qubit interactions are typically realized through exchange coupling between neighboring spins, coupling spin qubits to photons via hybrid spin-cQED devices enables long-range interactions and integration with other cQED platforms. Here, we investigate hole spin-photon coupling in compact single quantum dot setups. By incorporating ubiquitous strain inhomogeneities to our theory, we identify three main spin-photon coupling channels: a vector-potential-spin-orbit geometric mechanism--dominant for vertical magnetic fields--,  an inhomogeneous Rashba term generalizing previous spin-orbit field models, and strain-induced $g$-tensor terms--most relevant for in-plane fields. Comparing Si, unstrained (relaxed) Ge, and biaxially strained Ge wells, we find that Si and unstrained Ge provide optimal coupling strengths (tens of MHz) thanks to their reduced heavy-hole, light-hole splitting. We demonstrate efficient switching of the spin-photon coupling while preserving sweet spot operation. Finally, we evaluate quantum state transfer and two-qubit gate protocols, achieving $>99\%$ fidelity for state transfer and $>90\%$ for two-qubit gates with realistic coherence times, establishing single-dot hole spins as a viable platform for compact spin-cQED architectures and highlighting unstrained Ge as a promising candidate for spin-photon interactions.
\end{abstract}

\maketitle

\section{\label{sec:intro}Introduction}
Spin qubits in semiconductor quantum dots are a promising gate-tunable platform for quantum information processing~\cite{loss1998quantum, zwanenburg2013silicon, burkard2023semiconductor}. Hole spin qubits are especially attractive: their strong intrinsic spin-orbit coupling (SOC) enables ultrafast, all-electrical control~\cite{golovach2006electric, maurand2016cmos,watzinger2018germanium,crippa2018electrical,hendrickx2020single,froning2021ultrafast,jirovec2021singlet,wang2022ultrafast, fang2023recent, liles2024singlet}, and have demonstrated coherence times in the tens of microseconds~\cite{piot2022single, hendrickx2024sweet, carballido2024compromise, bassi2024optimal} alongside high-fidelity single- and two-qubit gates~\cite{hendrickx2020fast,geyer2024anisotropic}. Recent progress includes four- and ten-qubit Ge processors~\cite{hendrickx2021four,john2024two} and shared control across a 16-dot device~\cite{borsoi2024shared}.

Single-dot Loss-DiVincenzo spin qubits~\cite{loss1998quantum} natively couple to neighbors via exchange interaction, enabling fast and versatile two-qubit gates for both electron and hole spins~\cite{veldhorst2015two, he2019two, xue2019benchmarking, hendrickx2020fast, geyer2024anisotropic}. However, exchange is short-ranged, while fault-tolerant architectures require nonlocal connectivity~\cite{fowler2012surface, vandersypen2017interfacing}. Two main routes are  candidates to extend the range: coherent spin shuttling~\cite{sanchez2014long,mills2019shuttling,zwerver2023shuttling,van2024coherent,kunne2024spinbus,langrock2023blueprint,fernandez2024flying,ginzel2024scalable} and photon-mediated interactions in circuit QED (cQED)~\cite{blais2004cavity, blais2021circuit}, which also facilitate hybridization with other cQED-compatible platforms~\cite{scarlino2019coherent,landig2019virtual}.

In hybrid semiconductor-cQED devices, a cavity couples capacitively/galvanically to gates, introducing a quantized voltage into the semiconductor. Since spin degrees of freedom couple weakly to cavity photons~\cite{hu2012strong}, most original demonstrations first coupled photons to charge/orbital modes~\cite{mi2017strong, koski2020strong, kratochwil2021charge, de2024strong, janik2025strong} and then achieved coupling to spin via artificial (electrons)~\cite{samkharadze2018strong, mi2018coherent, landig2018coherent} or direct (holes)~\cite{yu2023strong, noirot2025coherence} SOC. With the use of micromagnets, electron spins have shown photon-mediated spin-spin interactions~\cite{borjans2020resonant, harvey2022coherent} and an iSWAP gate~\cite{dijkema2025cavity}; for hole spins, strong spin-photon coupling has been achieved~\cite{yu2023strong}, though two-qubit gates remain to be demonstrated.

To date, spin-photon coupling has largely relied on flopping-mode states delocalized over multiple dots to enhance the dipole interaction. Theory indicates that strong spin-photon coupling is possible even for a single dot~\cite{kloeffel2013circuit, michal2023tunable, bosco2022fully, prem2024longitudinal}, which is appealing for compact layouts and because single-dot hole spins typically exhibit longer coherence times than flopping modes~\cite{piot2022single, dijkema2025cavity, noirot2025coherence}. Experimentally, Ref.~\cite{yu2023strong} reported \(\sim\)1 MHz single-dot spin-photon coupling, but the large cavity decay rate precluded entering the strong-coupling regime.

\begin{figure*}
\includegraphics[width=2\columnwidth]{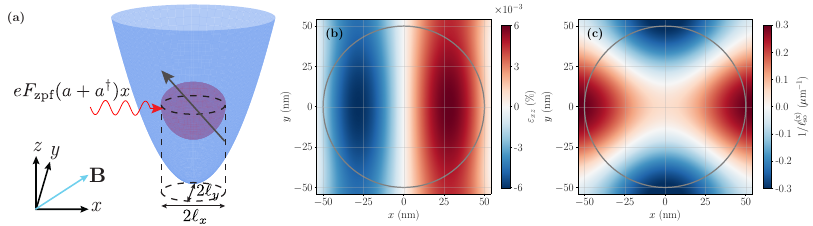}
\caption{Hybrid spin-photon system. (a) Anisotropic 2D harmonic quantum dot with lengths $\ell_{x,y}=\sqrt{\hbar/(m_\parallel\omega_{x,y})}$. The cavity field displaces the hole wave function along $x$ by $eF_\text{zpf}(a+a^\dagger)$, where $a^\dagger$ ($a$) creates (annihilates) a cavity photon. A magnetic field  $\mathbf{B}=B(\sin\theta\cos\phi,\sin\theta\sin\phi,\cos\theta)$ sets the Zeeman splitting of the ground-state spin doublet. 
(b) Shear-strain component $\varepsilon_{xz}(x,y)$ at $z=0$ for the strained Ge/GeSi device of Ref.~\onlinecite{abadillo2023hole} with a well thickness of 16 nm, a circular Al gate of 50~nm radius (grey circle) separated from the quantum well by 5~nm of insulating Al$_2$O$_3$ and 50~nm thick upper GeSi barrier. (c) Inverse inhomogeneous spin-orbit length $1/\ell_\text{so}^{(x)}(x,y)$ at $z=0$ for motion along $x$ in the same device.}
\label{fig:intro}
\end{figure*}

Hole-spin qubit control is highly susceptible to microscopic device details. Quantum dot geometry~\cite{michal2021longitudinal, bosco2021squeezed,martinez2022hole}, inhomogeneous strain~\cite{liles2021electrical, abadillo2023hole, wang2024electrical}, interface roughness~\cite{martinez2025variability}, and local material variations~\cite{rodriguez2023linear} all impact performance and variability. In particular, naturally arising strain fields due to the gate stack introduce spatial gradients of the \(\hat g\)-matrix and inhomogeneous SOC~\cite{abadillo2023hole}, strongly enhancing manipulability. 

Here we develop a comprehensive theory of single-dot hole spin-photon coupling that incorporates these effects in Si, biaxially strained Ge, and the recently proposed unstrained Ge heterostructures~\cite{costa2025buried, mauro2025hole}. We find new terms arising from  inhomogeneous strain fields which enable spin-photon couplings of a few tens of MHz, more effectively in unstrained Ge. We also analyze strategies to switch the interaction on and off and find that detuning the qubit from the cavity via gate-controlled confinement is the most practical and coherence-preserving approach.

The manuscript is organized as follows. 
Sec.~\ref{sec:effectivemodel} introduces the effective model including inhomogeneous spin-orbit mechanisms. 
Sec.~\ref{sec:spinphoton} evaluates spin-photon coupling across material platforms, benchmarking to tight-binding simulations, and finding unstrained Ge as the optimal material for spin-photon interactions. 
We then compare different coupling control strategies for the optimal unstrained Ge in Sec.~\ref{sec:tunability}. Finally, in Sec.~\ref{sec:twoqubit} we present resonant and dispersive two-qubit gate protocols and their fidelities in unstrained Ge devices. 
Conclusions are given in Sec.~\ref{sec:conclusion}.

\section{Model}
We begin by developing an effective model for a hole spin confined in a semiconductor nanostructure and its interaction with a cavity photon.
\label{sec:effectivemodel}
\subsection{2D model for a confined hole spin}
The heavy-hole (HH) and light-hole (LH) envelope functions obey the Hamiltonian:
\begin{equation}
H_\text{h}=H_\mathrm{K}+H_\mathrm{BP}+H_\mathrm{Z}+V(\mathbf{r})\mathds{1}\,,
\label{eq:H4KP}
\end{equation}
where $H_\mathrm{K}$ is the Kohn-Luttinger Hamiltonian \cite{luttinger1956quantum}, $H_\mathrm{BP}$ is the Bir-Pikus Hamiltonian describing the effects of strain~\cite{bir1963spin, bir1963spin2}, $H_\mathrm{Z}$ is the Zeeman term, and $V(\mathbf{r})$ is the confining electrostatic potential with $\mathds{1}$ as the identity matrix in the hole subspace. 

We assume strong vertical confinement along $z=[001]$, producing a HH-LH splitting $\Delta_\mathrm{LH}\approx \frac{2\pi^2\hbar^2\gamma_2}{m_0L_\mathrm{W}^2}+2b_v(\varepsilon_\parallel-\varepsilon_\perp)$, where $L_\mathrm{W}$ is the well thickness, $m_0$ the bare electron mass, $\gamma_i$ the Luttinger parameters, $b_v$ a deformation potential, and $\varepsilon_{xx}=\varepsilon_{yy}=\varepsilon_\parallel$, $\varepsilon_{zz}=\varepsilon_\perp$ the biaxial strains (see Table~\ref{tab:params}).

\begin{table}[h]
\centering
\begin{tabular}{l | r r r r r r r r}
\toprule
 & $\gamma_1$ & $\gamma_2$ & $\gamma_3$ & $a_v$ (eV) & $b_v$ (eV) & $d_v$ (eV) & $\kappa$ & $q$ \\
\hline
Si & 4.285 & 0.339 & 1.446 & 2.10 & $-2.330$ & $-4.750$ & $-0.420$ & 0.01 \\
Ge & 13.380 & 4.240 & 5.690 & 2.00 & $-2.160$ & $-6.060$ & 3.410 & 0.06 \\
\hline
\botrule
\end{tabular}
\caption{Luttinger parameters $\gamma_i$, deformation potentials $a_v, \, b_v,\, d_v $, and Zeeman parameters of Si and Ge at $T=0$\,K.~\cite{winkler2001spin}}
\label{tab:params}
\end{table}

Under vertical confinement, the ground state has predominantly HH character and an effective 2D theory is obtained by performing a Schrieffer-Wolff transformation integrating the vertical direction and the LH states in Eq.~\eqref{eq:H4KP}~\cite{winkler2001spin, kloeffel2011strong, marcellina2017spin, terrazos2021theory, michal2021longitudinal, wang2022ultrafast, abadillo2023hole}. In the effective 2D theory, the HH spin is described by
\begin{equation}
    H_\text{h}^\text{eff}=\frac{\Pi_x^2}{2m_\parallel}+\frac{\Pi_y^2}{2m_\parallel}+V_\text{2D}(x,y)+\frac{1}{2}\hbar\boldsymbol{\omega}_L(x,y)\cdot\boldsymbol{\sigma}+H_\text{soc},
    \label{eq:2dbase}
\end{equation}
where $\Pi_i=p_i+eA_i$ (with $A_i$ the $i$th component of the vector potential), $m_\parallel\approx m_0/(\gamma_1+2\gamma_2)$ is the in-plane HH effective mass, and the quantum dot potential is taken to be harmonic: $V_\text{2D}(x,y)=\tfrac{1}{2}m_\parallel(\omega_x^2x^2+\omega_y^2y^2)$; see Fig.~\ref{fig:intro}(a). The Larmor vector $\boldsymbol{\omega}_L$ may depend on the size and position in 2D space of the HH wavefunction, and is written as $\hbar\boldsymbol{\omega}_L=\mu_B\hat{g}(x,y)\mathbf{B}$ in terms of the magnetic field $\mathbf{B}=(B_x,B_y,B_z)$ and the position-dependent $\gt$-matrix (with $\mu_B$ the Bohr magneton). Here, $\boldsymbol{\sigma}$ acts on the HH ground-state Kramers doublet.

For holes, $\hat{g}$ heavily depends on confinement and strain. Shear strain, for instance, introduces off-diagonal components to the $\hat{g}$-matrix. Explicit expressions of the $\hat{g}$-matrix in terms of microscopic details such as strain and confinement are given in Appendix~\ref{app:2d}. As an example, we show in Fig.~\ref{fig:intro}(b) the shear-strain component $\varepsilon_{xz}$ as a function of position within a nanostructure with an Al gate of 50 nm radius. This shear strain component is proportional to the $\hat{g}$-matrix element $\hat{g}_{zx}$, introducing rotations to the axes of the $\hat{g}$-matrices. The $\hat{g}$-matrix encodes part of the spin-orbit effects, while $H_\text{soc}$ contains the kinetic SOC mechanisms, including  inhomogeneous SOC. 

The kinetic SOC of vertically confined HHs is a cubic-in-momentum Rashba interaction; however, for moderate anisotropy this contribution is typically weaker than the inhomogeneous Rashba SOC arising from strain gradients~\cite{abadillo2023hole}. Accordingly, we take $H_\text{soc}$ to be dominated by inhomogeneous SOC. For planar devices, the leading inhomogeneous Rashba term is
\begin{equation}
    H_\text{soc}\approx \frac{\hbar}{m_\parallel}\left(\Bigg\{\frac{1}{l_\text{so}^{\text{(x)}}(x,y)},\Pi_x\Bigg\}\sigma_y+\Bigg\{\frac{1}{l_\text{so}^{\text{(y)}}(x,y)},\Pi_y\Bigg\}\sigma_x\right),
    \label{eq:inhomosoc}
\end{equation}
where $\{A,B\}=(AB+BA)/2$, and the spin-orbit lengths $\ell_\text{so}^{(x/y)}$ can be related to strain and material parameters. For illustration, Fig.~\ref{fig:intro}(c) shows the inverse inhomogeneous spin-orbit length as a function of position within a nanostructure with an Al gate of 50 nm radius. We refer to Appendix~\ref{app:2d} for the explicit expressions of the spin-orbit length in terms of microscopic details. 

\subsection{Light-matter interaction}
We now include the cavity and its coupling to the dot. We consider a lateral gate coupling that induces a quantized in-plane electric field $F_\text{zpf}$ along $x$ (dipole gauge), which displaces the hole wave function; see Fig.~\ref{fig:intro}(a). Let $a^\dagger$ ($a$) create (annihilate) a cavity photon of frequency $\omega_R$. The hybrid Hamiltonian is 
\begin{equation}
    H_\text{spin-photon}=H_\text{h}^\text{eff}+\hbar\omega_Ra^\dagger a+eF_\text{zpf}x(a+a^\dagger).
    \label{eq:hybrid2D}
\end{equation}
To explore the dynamics within the lowest orbital state, we perform a Schrieffer-Wolff transformation~\cite{winkler2001spin} that eliminates excited orbitals and projects onto the subspace spanned by the spin states in the orbital Fock-Darwin ground state and arbitrary photon number. Details of such projection are given in Appendix~\ref{app:fockdarwin}. Keeping terms up to second order in the perturbative couplings (small $\delta x$ compared to $\ell_x$, and weak SOC gradients and Zeeman energy compared to the orbital spacings $\hbar\omega_{x,y}$), we obtain the effective spin-photon Hamiltonian:
\begin{equation}
    \begin{aligned}
        H_\text{spin-photon}^\text{eff}&=\hbar\omega_Ra^\dagger a+\frac{1}{2}\hbar\boldsymbol{\omega}_L\cdot\boldsymbol{\sigma}+H_\text{int} \\
        H_\text{int}&=H_\text{R}+H_\varepsilon+H_\text{v},
    \end{aligned}
    \label{eq:spinphoton}
\end{equation}
where the Larmor vector is projected onto the orbital ground state and small cavity-frequency renormalizations are neglected. The interaction splits into three physically distinct terms: $H_\text{R}$ (Rashba term, driven by the displacement along $x$), $H_\varepsilon$ (the strain-gradient induced $\hat g$-matrix modulation along $x$), and $H_\text{v}$ (a unique vector-potential/SOC/photon interference term). Defining $\delta x=eF_\text{zpf}/(m_\parallel\omega_x^2)$ as the field-induced displacement and $\mathbf{n}_\text{so}^{(x_i)}$ as the spin-orbit direction for motion along $x_i$, their general forms, for arbitrary strain
and SOC profiles, are
\begin{subequations}
\begin{align}
    H_\text{R}&=\hbar\bigg\langle\frac{\delta x}{2\ell_\text{so}^{(x)}}\bigg\rangle\Big[(a+a^\dagger)(\boldsymbol{\omega}_L\times\mathbf{n}^{(x)}_\text{so})\cdot \boldsymbol{\sigma}\nonumber \\ &-i\omega_R(a-a^\dagger)(\boldsymbol{\sigma}\cdot\mathbf{n}_\text{so}^{(x)})\Big] \\
    H_\varepsilon &= \frac{\delta x}{2}\mu_B\boldsymbol{\sigma}\cdot\left(\frac{\partial \hat{g}}{\partial x}\right)\cdot\mathbf{B}(a+a^\dagger) \\
    H_\text{v}&=\delta x \langle \partial_x H_\text{soc}\rangle(a+a^\dagger)\,,
\end{align}
\label{eq:genspinphoton}
\end{subequations} 
where $\langle\cdot\rangle$ denotes the expectation values in the orbital ground state. Note that $H_\text{v}$ is not proportional to the Larmor frequency and functions as an effective spin-orbit driving term. For concreteness, we next consider planar dots defined by quasi-circular gates. Following Ref.~\cite{abadillo2023hole}, we simulate the strain fields in such structures and obtain compact analytical fits for the spatial dependence of the $\gt$-matrix and the inhomogeneous SOC, summarized in Table~\ref{tab:compact_fits}. For these planar devices, the interaction terms simplify to 
\begin{subequations}
\begin{align}
    H_\text{R}&\approx \hbar\bigg\langle\frac{\delta x}{2\ell_\text{so}^{(x)}}\bigg\rangle \nonumber \\ &\times\Big[(a+a^\dagger)(\boldsymbol{\omega}_L\times\mathbf{e}_y)\cdot \boldsymbol{\sigma}-i\omega_R(a-a^\dagger)\sigma_{y}\Big] \\
    H_\varepsilon &\approx  -2\frac{\sqrt{3}\kappa d_v\delta x}{\Delta_\mathrm{LH}}\mu_BB_x\sigma_zp_{xz}(a+a^\dagger) \\
    H_\text{v}&\approx \frac{\delta x\ell_x^2\ell_y^2}{\ell_x^2+\ell_y^2}\beta_{yx}\hbar\omega_B(a+a^\dagger)\sigma_x, 
\end{align}
\label{eq:specificspinphoton}
\end{subequations} 
where $\mathbf{e}_y$ is the unit vector along $y$, $\omega_B=eB_z/m_\parallel$ is the cyclotron frequency, $\ell_{x,y}=\sqrt{\hbar/(m_\parallel\omega_{x,y})}$, $p_{xz}$ is the gradient of the shear strain component $\varepsilon_{xz}\approx p_{xz}x$ (see Fig.~\ref{fig:intro}(b)), and $\beta_{yx}$ is a parameter quantifying the dependence of the spin-orbit length $\ell_\text{so}^{(y)}$ with respect to $x$, which we obtain from the inverse spin-orbit-length fits (see $1/\ell_{\mathrm{so}}^{(y)}$ in Table~\ref{tab:compact_fits}, part (ii)). 

\begin{figure*}
\includegraphics[width=1\textwidth]{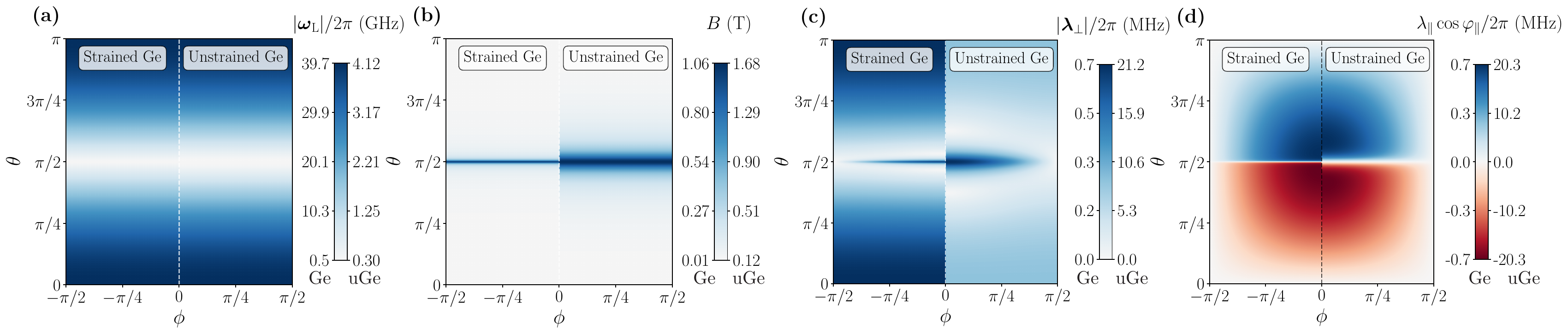}
\caption{Larmor frequency and spin-photon coupling for Ge with and without biaxial strain. The dot is circular with $\ell_x=\ell_y=30$\,nm and $F_\text{zpf}=30$\,V/m. We take $\Delta_\mathrm{LH}=70$\,meV (strained Ge) and $\Delta_\mathrm{LH}=3.5$\,meV (unstrained Ge)~\cite{costa2025buried}, and $\beta_{yx}^\text{(uGe)}\approx -4.1\times 10^{-6}$ nm$^{-3}$; $\beta_{yx}^\text{(Ge)}\approx -0.31\times 10^{-6}$ nm$^{-3}$; $p_{xz}^\text{(uGe)}\approx 3.9\times 10^{-6}$ nm$^{-1}$; $p_{xz}^\text{(Ge)}\approx 4.9\times 10^{-6}$ nm$^{-1}$; see Table~\ref{tab:compact_fits}. For each case, the plots are symmetric under $\phi\!\to\!-\phi$. (a) $|\boldsymbol{\omega}_L|/2\pi$ versus magnetic field orientation $\mathbf{B}=B(\sin\theta\cos\phi,\sin\theta\sin\phi,\cos\theta)$ at $B=0.2$\,T. (b) Magnetic field required to reach $|\boldsymbol{\omega}_L|/2\pi=2.5$\,GHz as a function of field orientation. (c) Transverse coupling $|\boldsymbol{\lambda}_\perp|/2\pi$ versus field orientation at $B=1$\,T. (d) Longitudinal coupling $\lambda_\parallel\cos\varphi_\parallel/2\pi$ versus field orientation at $B=1$\,T. Note the two different scales in the colorbar referring to the strained (left) and unstrained (right) cases.}
\label{fig:couplingsvsB}
\end{figure*}

$H_\text{R}$ is the spin-photon interaction term due to kinetic SOC, and characterizes the spin-photon interaction due to the spin-orbit field along the motion in $x$~\cite{bosco2022fully, michal2023tunable}. Our result is generalized for any spin-orbit length, including spatially inhomogeneous lengths, through the expectation value of $\ell_\text{so}^{(x)}$. The last two spin-photon interactions are obtained here for the first time: $H_\varepsilon$ is the quantized version of the strain-mediated $\hat g$-tensor modulation resonance incorporated through spatially-dependent $\hat{g}$-matrices, which dominate the AC hole spin driving for in-plane magnetic fields~~\cite{abadillo2023hole}. Strikingly, $H_\text{v}$ depends on the SOC along $y$ (orthogonal to the displacement due to the photon field), is proportional to $\omega_B$ rather than to $|\boldsymbol{\omega}_L|$, and can be understood as a geometric interaction due to the driving of the spin-orbit field in presence of a magnetic-field vector potential. Since the spin-orbit length along $y$ depends on $x$, by displacing the hole along $x$, the hole spin acquires a non-abelian phase. This term is particularly relevant for magnetic fields along the vertical direction due to their direct dependence on $\omega_B$.
 
\section{Spin-photon coupling}
\label{sec:spinphoton}
\subsection{General transverse and longitudinal interactions}

In general, the spin-photon interaction can rotate the spin about different axes depending on the underlying mechanism: whether the coupling is transverse or longitudinal is established by the relative orientation between the Larmor and  interaction vectors in spin space~\cite{michal2023tunable, bosco2022fully}. 
In the following, we analyze different scenarios and material platforms. We find that a magnetic field along the direction of motion optimizes transverse spin-photon interactions for a given Larmor frequency, while for a given magnetic field amplitude the vertical orientation leads to a larger coupling. Due to the smaller HH-LH splitting, these mechanisms are most efficient for unstrained Ge. 

To evaluate this,
it is useful to move to a frame in which the Zeeman term is diagonal and express the interaction relative to the Larmor vector. As an intermediate step, Eq.~\eqref{eq:spinphoton} can be grouped in terms of the field quadratures $a+a^\dagger$ and $-i(a-a^\dagger)$ as
\begin{equation}
    H=\hbar\omega_Ra^\dagger a+\frac{1}{2}\hbar\boldsymbol{\omega}_L\cdot\boldsymbol{\sigma}+\hbar\boldsymbol{\lambda}^{(X)}\cdot\boldsymbol{\sigma}(a+a^\dagger)-i\hbar\boldsymbol{\lambda}^{(P)}\cdot\boldsymbol{\sigma}(a-a^\dagger),
\end{equation}
where $\boldsymbol{\lambda}^{(X)}$ and $\boldsymbol{\lambda}^{(P)}$ are vectors in spin space, grouping the different terms in  Eqs.~(\ref{eq:genspinphoton}-\ref{eq:specificspinphoton}). After rotating to diagonalize the Larmor vector, we find
\begin{equation}
    \begin{aligned}
        \tilde{H}&=\hbar\omega_Ra^\dagger a+\frac{1}{2}\hbar|\boldsymbol{\omega}_L|\sigma_z+(ae^{-i\varphi_\parallel}+a^\dagger e^{i\varphi_\parallel})\hbar\lambda_\parallel\sigma_z\\ &+(ae^{-i\varphi_\perp}+a^\dagger e^{i\varphi_\perp})\hbar\boldsymbol{\lambda}_\perp\cdot\boldsymbol{\sigma},
    \end{aligned}
    \label{eq:finalspinphoton}
\end{equation}
which is an extended anisotropic Rabi model~\cite{xie2014anisotropic} including a longitudinal component~\cite{didier2015fast}. We define $\lambda_\parallel\cos\varphi_\parallel=(\boldsymbol{\omega}_L\cdot\boldsymbol{\lambda}^{(X)})/|\boldsymbol{\omega}_L|$ and $\lambda_\parallel\sin\varphi_\parallel=(\boldsymbol{\omega}_L\cdot\boldsymbol{\lambda}^{(P)})/|\boldsymbol{\omega}_L|$; similarly, $\boldsymbol{\lambda}_\perp\cos\varphi_\perp=(\boldsymbol{\omega}_L\times\boldsymbol{\lambda}^{(X)})/|\boldsymbol{\omega}_L|$ and $\boldsymbol{\lambda_\perp}\sin\varphi_\perp=(\boldsymbol{\omega}_L\times\boldsymbol{\lambda}^{(P)})/|\boldsymbol{\omega}_L|$. Physically, the longitudinal term produces a phase useful for fast readout~\cite{didier2015fast, harpt2025ultra, chessari2025unifying} and two-qubit phase gates~\cite{harvey2018coupling, bosco2022fully, michal2023tunable}. The transverse term enables dispersive readout and direct excitation exchange~\cite{blais2004cavity, blais2021circuit}.

To gain intuition into the relationship between the Larmor vector and the different couplings, we show how couplings and frequencies depend on magnetic field orientation  in Fig.~\ref{fig:couplingsvsB}. We focus on circular quantum dots and consider two types of Ge devices with different HH-LH splittings. The results for Si are qualitatively similar to those of unstrained Ge. On the one hand, we consider the usual biaxially strained Ge/Ge$_{0.8}$Si$_{0.2}$ quantum wells with a well depth of 16\,nm. The combination of both vertical confinement and biaxial strain leads to a large HH-LH splitting $\Delta_\mathrm{LH}\!\approx\!70$\,meV. On the other hand, we consider the recently developed Ge relaxed wells~\cite{stehouwer2025exploiting, costa2025buried}. These heterostructures are grown on Ge with a barrier of tensile strained Ge$_{0.8}$Si$_{0.2}$, leading to $\Delta_\mathrm{LH}\!\approx\!3.5$\,meV. We refer to these heterostructures as {\it unstrained} Ge wells, but note that inhomogeneous strains due to the gate stack also arise in such structures~\cite{mauro2025strain}. As a consequence of the different $\Delta_\mathrm{LH}$, the spin-orbit lengths and the strain-induced interaction in Eq.~\ref{eq:specificspinphoton} are roughly an order of magnitude larger for unstrained Ge than for strained Ge, such that $\beta_{yx}^\text{(Ge)}\approx -0.31\times 10^{-6}$ nm$^{-3}$ and $\beta_{yx}^\text{(uGe)}\approx -4.1\times 10^{-6}$ nm$^{-3}$; see Appendix~\ref{app:2d}.

\begin{figure*}
\includegraphics[width=2\columnwidth]{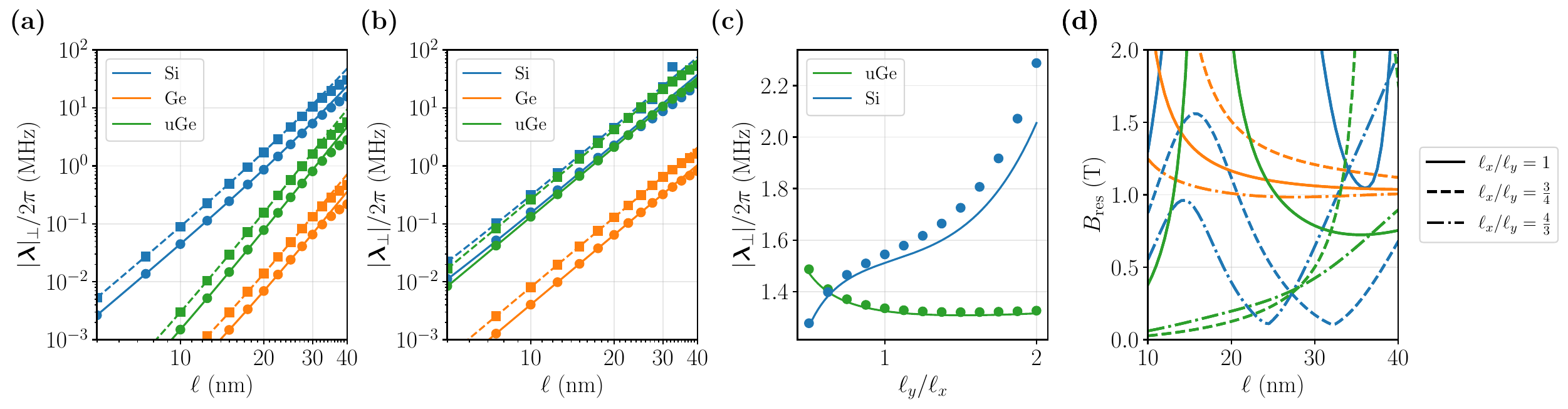}
\caption{Transverse spin-photon coupling and required magnetic field versus harmonic confinement. Parameters: $F_\text{zpf}=30$ V/m for Ge and uGe devices; $F_\text{zpf}=90$\,V/m\, since Si lever arms can be three times larger than in Ge~\cite{yu2023strong}$; \Delta_\mathrm{LH}=70$ meV (strained Ge) and $\Delta_\mathrm{LH}=3.5$ meV (unstrained Ge and Si); $\beta_{yx}^\text{(Si)}\approx -3.8\times 10^{-6}$ nm$^{-3}$; see Table~\ref{tab:compact_fits}; and we assume the resonance condition $\omega_R=|\boldsymbol{\omega}_L|$. (a) $\normgperp$ versus harmonic length for circular dots ($\ell_x=\ell_y=\ell$) at $B_z=0.1$ T (solid) and $0.2$ T (dashed), from the analytical formulas in Eq.~\ref{eq:specificspinphoton}. The data points show tight-binding simulations based on Eq.~\ref{eq:hybrid2D} (Appendix~\ref{app:TB}). (b) As in (a) but for in-plane fields $B_x=0.5$ T (solid) and $1$ T (dashed). (c) $\normgperp$ versus anisotropy ratio $\ell_y/\ell_x$ at fixed $\ell_x=15$ nm and $B_x=1$ T. (d) Magnetic field along $x$ required to reach $|\boldsymbol{\omega}_L|/2\pi=2.5$ GHz versus $\ell_x$ for Si (blue), Ge (orange), and unstrained Ge (green), for several anisotropy ratios $\ell_y/\ell_x$.}
\label{fig:couplingsvsL}
\end{figure*}

In Fig.~\ref{fig:couplingsvsB}(a), we show the Larmor frequency versus magnetic field orientation for $B=0.2$\,T, assuming a circular dot with $\ell_x=\ell_y=30$\,nm. We define $\theta$ and  $\phi$ such that $\mathbf{B}=B(\sin\theta\cos\phi,\sin\theta\sin\phi,\cos\theta)$. Complementarily, Fig.~\ref{fig:couplingsvsB}(b) shows the field required to reach $|\boldsymbol{\omega}_L|/2\pi=2.5$\,GHz. The large vertical $g$-factors of HHs ($g_{zz}^{(\mathrm{Ge})}\approx 14$ and $g_{zz}^{(\mathrm{uGe})}\approx 1.5$~\cite{mauro2025hole, costa2025buried}) make the Larmor frequency grow very fast with a vertical field, reaching 2.5\,GHz with $B_z\approx 0.013$\,T (strained Ge) and $B_z\approx 0.12$\,T (unstrained Ge). In contrast, in-plane HH $g$-factors are small in both cases, and fields of the order of 1\,T are required to reach a few GHz of Larmor frequency. Note that large fields degrade superconducting resonator quality factors $Q$, especially for out-of-plane orientations~\cite{yu2021magnetic}.

In Fig.~\ref{fig:couplingsvsB}(c), we plot $|\boldsymbol{\lambda}_\perp|/2\pi$ versus field orientation at $B=1$\,T with $F_\text{zpf}=30$\,V/m\ (Ge)~\footnote{The value $F_\text{zpf}=30$\,V/m corresponds to a voltage $V_\text{zpf}=20\,\mu$eV, achievable in Ge~\cite{janik2025strong}, across lateral gates separated by 133\,nm~\cite{abadillo2023hole} with lever arm 0.2~\cite{janik2025strong}. }. The pattern resembles classical-drive Rabi maps~\cite{martinez2022hole}: for displacement along $x$, a local maximum occurs for $\mathbf{B}\parallel x$ ($\theta=\frac{\pi}{2}$, $\phi=0$); another maximum in the coupling is found for $\mathbf{B}\parallel z$ ($\theta=0$). However, in between these two cases, for an intermediate orientation in the $x-z$ plane, the coupling tends to zero, indicating that the different contributions exactly cancel out. Moreover, the coupling exactly vanishes for $\mathbf{B}\parallel y$ ($\theta=\frac{\pi}{2}$, $\phi=\frac{\pi}{2}$). These trends follow from Eq.~\ref{eq:specificspinphoton}. The dominant terms are $H_\varepsilon$ for $\mathbf{B}\parallel x$ (\(\sim\!0.5\)\,MHz in strained Ge; \(\sim\!21\)\,MHz in unstrained Ge) and $H_\text{v}$ for $\mathbf{B}\parallel z$ (\(\sim\!0.7\)\,MHz in strained Ge; \(\sim\!5\)\,MHz in unstrained Ge). These contributions have opposite sign, resulting in the cancelation for intermediate angle $\theta$. The Rashba term $H_\text{R}$ also contributes to the spin-photon interaction for $\mathbf{B}$ in the $x-z$ plane, but its contribution is negligible compared to the other two terms.  

Strained Ge leads to much lower spin-photon couplings than unstrained Ge due to the differences in $\Delta_\mathrm{LH}$. Besides, the width of the spin-photon peak at $\mathbf{B}\parallel x$ is much wider for unstrained Ge than strained Ge due to the more isotropic $g$-factors of unstrained Ge. This is an advantage for unstrained Ge since it makes the spin properties more robust against magnetic field misalignment. Considering single-dot hole-spin coherence times are in the few to tens of microseconds~\cite{piot2022single, bassi2024optimal}, these spin-photon values approach the strong-coupling regime, especially for unstrained Ge.

Finally, the longitudinal coupling $\lambda_\parallel\cos\varphi_\parallel$ in Fig.~\ref{fig:couplingsvsB}(d) exhibits the opposite pattern to the transverse one, a geometric effect known as reciprocal sweetness~\cite{michal2021longitudinal} that swaps maxima and minima between transverse and longitudinal terms. The inclusion of the cosine term allows us to focus on the part that goes with the $a+a^\dagger$ quadrature, which is strictly proportional to the magnetic field rather than the resonator frequency. Maximal $\lambda_\parallel\cos\varphi_\parallel$ occurs for orientations between $x$ and $z$, with a symmetry-enforced zero for in-plane magnetic fields. The peak values ($\lambda_\parallel/2\pi\approx0.7$\,MHz in strained Ge; $\approx20$\,MHz in unstrained Ge) are comparable to the transverse case and likewise dominated by strain and vector-potential terms. While longitudinal coupling is not our focus, it tracks dephasing from charge noise due to electric field fluctuations along $x$, motivating operation where $\lambda_\parallel\cos\varphi_\parallel$ is minimal.

\subsection{Size dependence}

The interaction terms in Eqs.~\eqref{eq:specificspinphoton} have different dependence on wavefunction shapes and strain profiles. In particular, the harmonic lengths $\ell_x$ and $\ell_y$ are gate-tunable and strongly impact the coupling, with an increase  wavefunction size leading to an overall enhanced coupling strength. While Fig.~\ref{fig:couplingsvsB} fixed $\ell_x=\ell_y=30$\,nm, the overall length dependence is shown in Fig.~\ref{fig:couplingsvsL}, for Si, strained and unstrained Ge, under the resonant condition $\omega_R=|\boldsymbol{\omega}_L|$. In Fig.~\ref{fig:couplingsvsL}(a), we evaluate $\lambda_\perp$ for a purely vertical field $B_z$ assuming circular dots ($\ell_x=\ell_y=\ell$). We show results from Eqs.~(\ref{eq:specificspinphoton}-\ref{eq:finalspinphoton}) and from numerical tight-binding calculations; see Appendix~\ref{app:TB}. The coupling grows rapidly with size for vertical fields, with $\propto \ell^6$ (Ge) and $\propto \ell^4$ (Si) scalings. For $B_z=0.1$\,T, it exceeds 1\,MHz for $\ell\!\approx\!20$\,nm (Si) and $\ell\!\approx\!30$\,nm (unstrained Ge); strained Ge would require $\gtrsim\!40$\,nm, which is harder to realize. The power laws can be traced to the underlying mechanisms: for vertical fields, the vector-potential term in Eq.~\eqref{eq:specificspinphoton}(c) yields a prefactor $\delta x=eF_\text{zpf}m_\parallel \ell_x^4/\hbar^2$ (quartic in $\ell$) multiplied by $\ell_x^2/2$ for circular dots, giving the $\ell^6$ trend in Ge\,\footnote{We ignore here any residual $\ell_x$ dependence of $\beta_{yx}$, a device parameter that depends on gate geometry rather than wavefunction size.}. The different Si scaling indicates Rashba dominance (approximately $\propto\ell_x^4$ via $\delta x$), consistent with the larger Rashba coefficient in Table~\ref{tab:compact_fits}. Analytical and tight-binding results, shown as discrete data points, agree well; small deviations appear near $\ell_x\!\approx\!35$\,nm due to nearby excited states, marking the limits of our perturbation theory. Note that Si hole wavefunctions are typically smaller ($5-15$ nm) than in Ge due to the heavier mass, hence the largest sizes considered in Fig.~\ref{fig:couplingsvsL} are not relevant for Si.

Figure~\ref{fig:couplingsvsL}(b) shows the length dependence for in-plane fields parallel to the cavity drive, $B_x=0.5$\,T (solid) and $1$\,T (dashed). We observe $\normgperp\propto \ell_x^4$, as expected from the strain-driven term in Eq.~\eqref{eq:specificspinphoton}(b) via $\delta x$. In Si, Rashba contributes significantly, yet its scaling is indistinguishable from the strain-driven mechanism. For $B_x=1$\,T, couplings $\gtrsim$\,MHz are feasible for $\ell\!\approx\!15$\,nm in Si and unstrained Ge; strained Ge typically requires $\gtrsim30$\,nm, more accessible than in the vertical-field case.

In Fig.~\ref{fig:couplingsvsL}(c) we elongate the wavefunction by varying $\ell_y/\ell_x$ while keeping $\ell_x=15$\,nm and $B_x=1$\,T. Since $\delta x$ stays constant, this isolates other contributions to the scaling. Unstrained Ge shows a weak anisotropy dependence, whereas Si shows a stronger one, consistent with stronger Rashba dominance in Si. The expectation value of the inhomogeneous Rashba field has the form $\langle 1/\ell_\text{so}^{(x)}\rangle\approx\beta_0+\beta_{xx}\ell_x^2/2+\beta_{xy}\ell_y^2/2$ with $\beta_{xx}\approx-\beta_{xy}$ (Fig.~\ref{fig:intro}(c) and Table~\ref{tab:compact_fits}). For isotropic dots the extra length dependence mostly cancels; but anisotropy unbalances it, enhancing the Rashba effect in Si. Strong anisotropy can also amplify cubic Rashba SOC~\cite{bosco2021squeezed}, but it remains subdominant to inhomogeneous Rashba.

Comparing length trends and materials, the small HH-LH splitting makes Si and unstrained Ge optimal for large single-dot couplings. Among them, larger values are more accessible in unstrained Ge, where the lighter mass allows larger dots. Per unit applied magnetic field, vertical fields are optimal; however, the very large vertical $g$-factors make practical operation in the microwave resonator regime challenging: reaching a few GHz of Larmor frequency for cQED requires only mT fields, which would in turn yield very small couplings at resonance. In contrast, operation with in-plane fields ($B_x\!\sim\!0.5$--2\,T) supports sizable couplings while keeping the qubit in the GHz range. Figure~\ref{fig:couplingsvsL}(d) shows $B_\text{res}$, defined as the required $B_x$ field to reach the resonance condition $\qfreq=\omega_R/2\pi=2.5$\,GHz versus $\ell_x$ for different anisotropies. Across most cases shown, the required fields still yield MHz-level couplings. Strained Ge offers the least tunability, consistent with its larger HH-LH splitting, with Si and unstrained Ge displaying great Larmor frequency tunability. Overall, in-plane fields appear optimal for both coupling and ensuring GHz-range operation.

\section{Switching the spin-photon interaction}
\label{sec:tunability}
\begin{figure*}
\includegraphics[width=2\columnwidth]{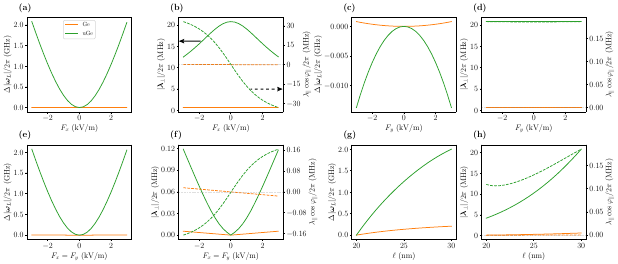}
\caption{Gate control of $|\boldsymbol{\omega}_L|$ and spin-photon couplings in planar devices for circular dots and with parameters relevant for Ge. In all panels: $\ell_x=\ell_y=30$\,nm and $F_\text{zpf}=30$\,V/m; $B_x=1$~T except in (e) and (f), where $B_y=1$~T. (a) Change in Larmor frequency $\Delta\qfreq$ versus in-plane electric field $F_x$. (b) $\normgperp$ (solid, left axis) and $\lambda_\parallel/2\pi$ (dashed, right axis) versus $F_x$. (c,d) Same as (a,b) but versus $F_y$. (e,f) Same as (a,b) for diagonal bias $F_x=F_y$ with $B_y=1$~T. (g,h) Same as (a,b) but versus harmonic length.}
\label{fig:tunability}
\end{figure*}

In the previous section, we showed that spin-photon coupling is most practical for a magnetic field applied parallel to the cavity drive, and that unstrained Ge is the optimal material for nearly isotropic dots. Hence, we focus on the in-plane operation of unstrained Ge in the following. In order to control the spin manipulation, it is desirable to switch the interaction on and off to toggle between idle (decoupled) and interacting regimes. In general, the coupling can be turned off either directly by driving $\normgperp\to 0$, or indirectly by detuning the qubit far from the cavity into the dispersive regime. We explore both methods here under the constraints of planar quantum-dot devices. In a planar device, it is possible to tune the spin-photon interaction by displacing the wavefunctions with lateral gates through electric fields $F_x$ and $F_y$ and modifying the harmonic confinement $\ell_x$ and $\ell_y$ by involving the plunger gate.

Figure~\ref{fig:tunability} compares several tuning strategies. We set a circular dot with $\ell_x=\ell_y=30$\,nm, a reasonable size for Ge dots, and use $B_x=1$\,T for panels (a)-(d) and (g)-(h), and $B_y=1$\,T for (e)-(f). Since the dot is circular, the unbiased Larmor frequency is the same across panels, $\qfreq\approx 3.2$\,GHz. Panels (a), (c), (e), and (g) show $\Delta\qfreq$ versus $F_x$, $F_y$, $F_x=F_y$, and $\ell$ ($\ell_x=\ell_y=\ell$), respectively. To illustrate the tunability of unstrained Ge, we include strained Ge for comparison. For $\mathbf{B}\parallel x$, the Larmor frequency is at a sweet spot with respect to both $F_x$ and $F_y$ at zero bias, so very small in-plane fields barely tune it. A field $F_x\approx 3$\,kV/m corresponds to a displacement $\delta x\approx 0.0086\,\ell_x$ and to a gate-voltage difference of 2\,mV for 130\,nm gate spacing with lever arm 0.2. Tuning with $F_x$ is very effective (roughly 2\,GHz over the shown range), while $F_y$ tuning, in contrast, is inefficient. Physically, strain-induced $\hat g$-matrix rotation mixes $x$ and $z$ effective $g$ factors, increasing the effective $g$ when displacing along $x$. Thus, displacement along $x$ efficiently detunes the qubit away from the cavity frequency. Alternatively, changing the plunger gate (modifying $\ell$) is also effective: Fig.~\ref{fig:tunability}(g) shows that $\ell_x:30\,\mathrm{nm}\to29\,\mathrm{nm}$ yields $\Delta\qfreq\approx200$\,MHz (about $10\times$ the spin-photon coupling). Anisotropic length tuning behaves similarly. The dominant mechanism is the intrinsic length dependence of the effective $\hat g$-matrix (Appendix~\ref{app:2d}). Owing to its smaller HH-LH splitting, unstrained Ge is far more tunable than strained Ge in all metrics.

To assess direct on/off control, panels (b), (d), (f), and (h) show $\normgperp$ for the same four strategies. Varying $F_x$ modulates the coupling but not enough to switch it fully off; varying $F_y$ leaves it nearly constant. For the diagonal bias $F_x=F_y$ we consider $B_y$ (instead of $B_x$) to start from a point where the coupling is off by symmetry and attempt to turn it on--this increases the coupling, but not to a sufficiently large value to achieve the strong coupling regime. Finally, varying $\ell$ in Fig.~\ref{fig:tunability}(h) yields the strongest modulation; however, fully turning off the coupling would require a large change in length. In practice, it is more effective to detune away from the cavity frequency via $\ell_x$ changes than to force $\normgperp\to0$. 

Alongside the transverse coupling, we plot $\mathrm{Re}(\lambda_\parallel e^{i\varphi_\parallel})=\lambda_\parallel\cos\varphi_\parallel$, which equals the Larmor-frequency shift from an electric-field fluctuation $\delta F=F_\text{zpf}$. This quantifies susceptibility to charge noise. In Fig.~\ref{fig:tunability}(b), the qubit sits at a sweet spot at $F_x=0$ for $\mathbf{B}\parallel x$, and is largely insensitive to $F_y$ in Fig.~\ref{fig:tunability}(d) as well. The longitudinal coupling is negligible $\lambda_\parallel\cos\varphi_\parallel\approx 0$ and barely changes with $\ell$ as shown in Fig.~\ref{fig:tunability}(h). 
This metric helps identify the optimal strategy: the qubit can be efficiently tuned away from the cavity frequency both by changing $F_x$ or $\ell_x$, but $F_x\neq0$ displaces the operating point away from the sweet spot, while changing $\ell_x$ preserves it. Therefore, in terms of coherence properties, the preferred on/off protocol is the harmonic-length tuning, which keeps the hole spin at the sweet spot.

\section{Quantum state transfer and two-qubit gates}
\label{sec:twoqubit}
To evaluate the potential of the hybrid spin-cQED system in the single-dot regime, we study quantum state transfer between the qubit and the photon~\cite{blais2004cavity}, a sideband-based two-qubit gate protocol~\cite{abadillo2021long}, and an off-resonant dispersive two-qubit gate~\cite{blais2004cavity}. We focus now on unstrained Ge devices since they are the most promising for both single-dot spin-photon coupling and tunability. From now on, we choose the cavity frequency $\omega_R/2\pi=3$ GHz for all protocols.

\subsection{Quantum state transfer}
When the qubit and cavity are in resonance $|\boldsymbol{\omega}_L|=\omega_R$, the spin undergoes a coherent exchange with the photon, preserving the total number of excitations within the rotating-wave approximation (RWA). This enables translating spin quantum information into photonic quantum information, which can be used to interface with distant cQED-compatible systems. 

\begin{figure}
\includegraphics[width=1\columnwidth]{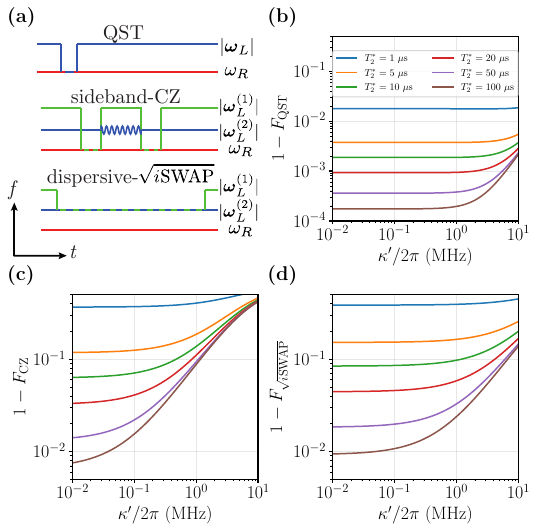}
\caption{Quantum state transfer and two-qubit gate protocols. (a) Pulse sequences: QST between qubit and photon (top), sideband-based CZ (middle), and dispersive $\sqrt{i\text{SWAP}}$ (bottom). Time and frequency are schematic. (b) QST average infidelity versus cavity decay rate $\kappa'$ for different $T_2^*$. (c) CZ average infidelity versus $\kappa'$ for the same $T_2^*$. (d) $\sqrt{i\text{SWAP}}$ average infidelity versus $\kappa'$ for the same $T_2^*$.}
\label{fig:twoqubit}
\end{figure}

The pulse sequence for performing this quantum state transfer (QST) is shown in Fig.~\ref{fig:twoqubit}(a). As the initial working point, we take $B_x=1$\,T and $\ell_x=\ell_y=30$\,nm, yielding $|\boldsymbol{\lambda}_\perp|/2\pi\approx21$\,MHz and $|\boldsymbol{\omega}_L|/2\pi\approx3.21$\,GHz, with the qubit at a sweet spot against small in-plane electric field fluctuations. The qubit and cavity are thus detuned by $|\boldsymbol{\omega}_L|-\omega_R\approx10|\gperp|$, i.e., in the dispersive regime. To activate the interaction, a plunger pulse that changes the harmonic length by $\delta\ell_x\approx-1$\,nm suffices to bring the qubit into resonance with the cavity, $\qfreq\approx3$\,GHz. At resonance, the effective Hamiltonian in the rotating frame within the RWA is
\begin{equation}
    H_\text{QST}\approx\hbar|\boldsymbol{\lambda}_\perp|(a\sigma_++a^\dagger\sigma_-),
    \label{eq:qst}
\end{equation}
which leads to coherent exchange between photon and qubit. 

Given the resonant nature of this protocol, QST fidelity is limited by both qubit decoherence and cavity decay. Hole-spin dephasing (typically limiting the spin decoherence process) is in the few to tens of $\mu$s~\cite{piot2022single, hendrickx2024sweet, bassi2024optimal}. We simulate QST with a Lindblad master equation in QuTIP~\cite{johansson2012qutip} including cavity decay $\kappa'$ and qubit decoherence $\gamma_2=2\pi/T_2^*$. The average gate fidelity is then computed via the Choi-Jamiolkowski isomorphism~\cite{gilchrist2005distance}. Figure~\ref{fig:twoqubit}(b) shows infidelity versus $\kappa'$ for several $T_2^*$. For this protocol, infidelities below $10^{-2}$ are obtained for $T_2^*$ in the few-$\mu$s range. The gate is extremely fast, with $\tau=\pi/(2|\gperp|)\approx12$\,ns, which enables high QST operation fidelities. As a resonant protocol, performance degrades with increasing $\kappa'$, especially for $\kappa'/2\pi\gtrsim1$\,MHz. These results confirm that high-fidelity QST is feasible with single-dot hole spins~\cite{michal2023tunable}.

\subsection{Sideband-mediated CZ gate}
A quasi-resonant photon-mediated CZ can be realized using sideband transitions together with near-resonant interactions~\cite{abadillo2021long}. Several sideband implementations exist for semiconductor qubits~\cite{srinivasa2016entangling,srinivasa2024cavity}; we adopt the approach of Ref.~\cite{strand2013first}, which requires the ability to drive the qubit frequency and is, therefore, well suited to our setup. 

Let us consider the interaction between a single qubit and the resonator together with a driving pulse on the qubit energy at our working point ($\lambda_\parallel=0$).
\begin{equation}
    \begin{aligned}
        \tilde{H}_\text{sb}&=\hbar\omega_Ra^\dagger a+\frac{1}{2}|\hbar\boldsymbol{\omega}_L|\sigma_z+\hbar(a+a^\dagger )|\boldsymbol{\lambda}_\perp|\sigma_x\\ &+\hbar\delta\omega\cos(\omega_dt+\varphi)\sigma_z,
    \end{aligned}
\end{equation}
By going to the frame $U(t)=\exp\left[-i\left(|\boldsymbol{\omega}_L|t+\frac{2\delta\omega}{\omega_d}\sin(\omega_dt+\varphi)\right)\sigma_z/2-i\omega_R a^\dagger a\right]$, we eliminate the driving term and the effective Hamiltonian in the RWA becomes
\begin{equation}
    \begin{aligned}
        H_\text{sb}\approx\hbar|\gperp|a^\dagger\sigma_-\sum\limits_{n=-\infty}^\infty J_n\left(\frac{2\delta\omega}{\omega_d}\right)e^{i\left[(\Delta-n\omega_d)t-\varphi\right]}+\text{h.c.},
    \end{aligned}
\end{equation}
where $J_n(x)$ are Bessel functions of the first kind, and $\Delta=|\boldsymbol{\omega}_L|-\omega_R$. For off-resonant operation, the largest coupling occurs for $n=1$, $\omega_d=\Delta$, and $\delta\omega\approx0.94\Delta$. For $\Delta/2\pi=0.21$\,GHz (same working point as QST), this entails $\delta\omega/2\pi\approx 197$\,MHz, achievable with $\delta\ell_x\approx-1$\,nm. In this regime, the red-sideband Hamiltonian in the RWA yields the unitary 
\begin{equation}
    S_-(|\gperp|t,\varphi)\approx\exp\left[-i|\gperp|t(e^{-i\varphi}a^\dagger\sigma_-+e^{i\varphi}a\sigma_+)/2\right].
\end{equation}
With these ingredients, a photon-mediated two-qubit gate can be constructed. For two qubits with frequencies $|\boldsymbol{\omega}_L^{(1)}|$ and $|\boldsymbol{\omega}_L^{(2)}|$, a CZ gate is implemented by the sequence~\cite{abadillo2021long}:
\begin{equation}
\begin{aligned}
U_\text{CZ}&=Z^{(2)}\left(-\frac{\pi}{\sqrt{2}}\right)Z^{(1)}\left(\frac{\pi}{\sqrt{2}}+\pi\right)S_-^{(1)}(\pi,0)\\ &\times S_-^{(2)}(\pi/2,0)S_-^{(2)}(\pi\sqrt{2},\pi/2)S_-^{(2)}(\pi/2,\pi)S_-^{(1)}(\pi,0),
\end{aligned}
\end{equation}
where the superscript denotes the target qubit and $Z$ is a $z$ rotation. The $\varphi=0$ sidebands used on qubit 1 can be implemented as QST segments without explicit AC driving. Figure~\ref{fig:twoqubit}(a) schematically depicts the CZ protocol. To mitigate residual dispersive interactions, it is helpful to keep $|\boldsymbol{\omega}_L^{(1)}|\neq|\boldsymbol{\omega}_L^{(2)}|$.

Analogously to QST, we simulate the sideband-mediated CZ with cavity decay $\kappa'$ and qubit dephasing $\gamma_2$ for the two qubits. We drive qubit 2 with $\delta\omega\approx100$\,MHz (optimizing the sideband), and realize the $\varphi=0$ segments on qubit 1 via QST. Figure~\ref{fig:twoqubit}(c) shows average CZ infidelity versus $\kappa'$ for various $T_2^*$. The CZ is longer than QST, with $\tau\approx9.7/|\boldsymbol{\lambda}_\perp|\approx73.5$\,ns for our parameters, and dephasing on both qubits reduces fidelity relative to QST. The $10^{-2}$ infidelity threshold is reached for $T_2^*\approx80$\,$\mu$s when $\kappa'$ is very small. Because this gate populates the cavity at different stages, it is partially sensitive to $\kappa'$. For $T_2^*$ values in the tens of microseconds, fidelities can exceed 90\% for low $\kappa'$. Larger $|\boldsymbol{\lambda}_\perp|$ (e.g., with high-impedance resonators) could enable higher fidelities.

\subsection{Dispersive gate}
Given that the CZ gate is partially limited by the cavity decay rate $\kappa'$, we consider a purely dispersive two-qubit gate. In the dispersive case, both qubits are sufficiently detuned from the cavity, and optimal operation uses two qubits at the same frequency. In the rotating frame $U(t)=\exp\left[-it\big(\sum\limits_{j=1,2}|\boldsymbol{\omega}_L|\sigma_z^{(j)}/2+\omega_Ra^\dagger a\big)\right]$, the interaction is~\cite{blais2021circuit}
\begin{equation}
    H_\text{dispersive}\approx \frac{|\gperp|^2}{\Delta}(\sigma_+^{(1)}\sigma_-^{(2)}+\sigma_-^{(1)}\sigma_+^{(2)}),
\end{equation}
where we assume identical spin-photon couplings for the two qubits. This interaction yields a $\sqrt{i\text{SWAP}}$ after $\tau=\frac{\pi\Delta}{4|\gperp|^2}$. Taking $\Delta\approx10|\gperp|$ and the same coupling as above ($\normgperp=21$\,MHz) gives $\tau\approx59.5$\,ns, comparable to the sideband gate.

Figure~\ref{fig:twoqubit}(d) shows simulated dispersive-gate fidelities using the same procedure as for QST and CZ. The $10^{-2}$ infidelity threshold is reached only for very large values of the coherence times $T_2^*\gtrsim100$\,$\mu$s at low $\kappa'$. For small $\kappa'$, the fidelity is slightly below the CZ case; however, because the dispersive gate does not populate the cavity, it is much less sensitive to $\kappa'$, and thus surpasses CZ at realistic cavity decay rates. As with CZ, for $T_2^*$ in the tens of microseconds, infidelities below 10\% are feasible with $\kappa'/2\pi$ in the MHz range; improved resonator impedance should further boost performance but fidelities above 99$\%$ may be very hard to achieve.

\section{Conclusion}
\label{sec:conclusion}

By explicitly including strain-induced spin-orbit effects, we show sizable and gate-tunable spin-photon interaction is possible for holes in single quantum dots.
We have gone beyond previously existing theory, finding that the spin-photon interaction is best described by an anisotropic Rabi model with longitudinal and transverse coupling and phases that depend on the magnetic field orientation. We show that spin-photon coupling is predominantly set by strain-driven $\hat g$-matrix variations for in-plane magnetic fields, whereas for out-of-plane fields it is dominated by a novel interference between the magnetic vector potential and inhomogeneous SOC. Couplings on the order of a few tens of MHz are achievable for hole-spin qubits with harmonic sizes of $\sim$25\,nm or larger, opening a realistic path to strong spin-photon coupling in near-term experiments.

Comparing the length dependence across materials, we find that large couplings are unattainable in biaxially strained Ge due to its large HH-LH splitting, which pushes the required dot sizes to impractical values. Si and unstrained Ge are better choices due to their reduced HH-LH splitting. In Si, the heavier mass limits realistic sizes to $\sim$5-15\,nm, yielding couplings of only a few MHz --consistent with observations~\cite{yu2023strong}. Unstrained Ge supports larger dots and, consequently, stronger couplings (a few tens of MHz). Moreover, the smaller HH-LH splitting in both Si and unstrained Ge greatly enhances gate tunability of coupling and qubit frequency, enabling interaction switching on/off by detuning via harmonic-length control.

The combination of relatively large couplings and high tunability in unstrained Ge enables efficient Quantum State Transfer and two-qubit gate protocols. QST benefits from short gate times and sweet-spot operation, supporting fidelities above the 99$\%$ threshold. The two-qubit gates considered here are slower and thus have lower fidelities, but values above 90\% are attainable with realistic cavity decay and $T_2^*$ in the few tens of microseconds, placing them within reach. Continued improvements in resonator impedance and $Q$, together with longer coherence, could enable high-fidelity two-qubit operations. Furthermore, it can be argued that the sizable spin-photon interaction may allow hole spin readout in compact (single-dot) setups, not requiring extra quantum dots as usual with Pauli spin blockade mechanisms. Our work highlights unstrained Ge as a very promising material platform for achieving strong spin-photon coupling and develop gate-based protocols for photon-mediated operations.

\acknowledgments
Work funded by MICIU/AEI/10.13039/501100011033, and European Union Next Generation EU/PRTR through Grants PID2021-125343NB-I00, RYC2022-037527-I, and PID2023-148257NA-I00. We also acknowledge the support of the Quantum Technologies Platform (QTEP-CSIC) and of the Severo Ochoa Centres of Excellence program through Grant CEX2024-001445-S.

\appendix

\setcounter{figure}{0}
\setcounter{equation}{0}
\setcounter{table}{0}
\renewcommand{\thefigure}{\thesection.\arabic{figure}}
\renewcommand{\theequation}{\thesection.\arabic{equation}}
\renewcommand{\thetable}{\thesection.\arabic{table}}

\section{Effective 2D model: $\gt$-matrix, and inhomogeneous spin-orbit interactions}
\label{app:2d}
In this Appendix, we relate the effective 2D Hamiltonian in Eq.~\eqref{eq:2dbase} to microscopic details such as the hole shape and the strain pattern. We recall the effective 2D Hamiltonian
\begin{equation}
    H_\text{h}^\text{eff.}=\frac{\Pi_x^2}{2m_\parallel}+\frac{\Pi_y^2}{2m_\parallel}+V_\text{2D}(x,y)+\frac{1}{2}\hbar\boldsymbol{\omega}_L(x,y)\cdot\boldsymbol{\sigma}+H_\text{soc}.
\end{equation}
We can split the Larmor vector contribution $\hbar\boldsymbol{\omega}_L=\mu_B\hat{g}\mathbf{B}$ in terms of two $\gt$-matrix contributions as $\hat{g}=\hat{g}^{(0)}+\delta\hat{g}$, where $\hat{g}^{(0)}$ are the bare $g$-factors, and $\delta\hat{g}$ are contributions from the strain field. The non-zero bare $\gt$-matrix elements are given by \cite{michal2021longitudinal, martinez2022hole}:
\begin{subequations}
\label{eq:gHH}
\begin{align}
\gt_{xx}^{(0)}&\approx+3q+\frac{6}{m_0\Delta_\mathrm{LH}}\left(\lambda \Pi_x^2-\lambda' \Pi_y^2\right) \\
\gt_{yy}^{(0)}&\approx-3q-\frac{6}{m_0\Delta_\mathrm{LH}}\left(\lambda \Pi_y^2-\lambda^\prime \Pi_x^2\right) \\
\gt_{zz}^{(0)}&\approx6\kappa+\frac{27}{2}q-2\gamma_h\,,
\end{align}
\label{eq:g}
\end{subequations}
where $\kappa$ and $q$ are given in Table~\ref{tab:params}$, \lambda=\kappa\gamma_2-2\eta_h\gamma_3^2$, $\lambda^\prime=\kappa\gamma_2-2\eta_h\gamma_2\gamma_3$, with $\gamma_h$ and $\eta_h$ depending on the material choice and confinement potential~\cite{ares2013nature,michal2021longitudinal}. Explicitly, we take $\gamma_h^{(Ge)}=3.56$, $\gamma_h^{(uGe)}=9.9$, and $\gamma_h^{(Si)}=1.16$; and $\eta_h^{(Ge)}=0.2$, $\eta_h^{(Ge)}=0.06$, and $\eta_h^{(Si)}=0.08$~\cite{michal2021longitudinal, mauro2025strain}. Neglecting higher-order magnetic field corrections, the expectation values of $\Pi_x$ and $\Pi_y$ for the ground-state HH envelope yield $\langle \Pi_i^2\rangle\approx-\hbar^2/\ell_i^2$, with harmonic lengths $\ell_i=\sqrt{\hbar/(m_\parallel\omega_i)}$. 

The natural presence of strain components $\varepsilon_{ij}(x,y)$ in the nanostructure introduces further corrections to the $\gt$-matrix~\cite{abadillo2023hole, liles2021electrical, piot2022single}, resulting in:
\begin{subequations}
\label{eq:deltag1}
\begin{align}
\delta\gt_{xx}&=\delta\gt_{yy}=\frac{6b_v\kappa}{\Delta_\mathrm{LH}}\langle\varepsilon_{yy}(x,y)-\varepsilon_{xx}(x,y)\rangle\\
\delta\gt_{zy}&=-\frac{4\sqrt{3}\kappa d_v}{\Delta_\mathrm{LH}}\langle\varepsilon_{yz}(x,y)\rangle \\
\delta\gt_{zx}&=-\frac{4\sqrt{3}\kappa d_v}{\Delta_\mathrm{LH}}\langle\varepsilon_{xz}(x,y)\rangle \\
\delta\gt_{xy}&=-\delta\gt_{yx}=\frac{4\sqrt{3}d_v\kappa}{\Delta_\mathrm{LH}}\langle\varepsilon_{xy}(x,y)\rangle\,.
\end{align}
\label{eq:deltag}
\end{subequations}
showing that shear strain components rotate the principal axes of the $\hat{g}$-matrix and that strain inhomogeneities translate into a spatial dependence of the $\hat g$-matrix elements. For illustration, Fig.~\ref{fig:intro}(b) shows the spatial dependence of $\varepsilon_{xz}(x,y,z=0)$ for the nanostructure simulated in Ref.~\cite{abadillo2023hole} with a 50\,nm-radius Al plunger gate. The Al gate induces a shear-strain gradient along $x$, yielding corrections $\delta\gt_{zx}(x,y)$ when the hole wavefunction is displaced along $x$.

The other spin-orbit term in the effective Hamiltonian is $H_\text{soc}$. For the devices considered here, an inhomogeneous Rashba interaction dominates the kinetic SOC. As mentioned in the main text, this leads to an effective Rashba of the form
\begin{equation}
    H_\text{soc}\approx \frac{\hbar}{m_\parallel}\left(\Bigg\{\frac{1}{l_\text{so}^{\text{(x)}}(x,y)},\Pi_x\Bigg\}\sigma_y+\Bigg\{\frac{1}{l_\text{so}^{\text{(y)}}(x,y)},\Pi_y\Bigg\}\sigma_x\right).
    \label{eq:inhomosocA}
\end{equation}
The spin-orbit lengths can be related to microscopic details through the formulas:
\begin{equation}
\begin{aligned}
    \frac{1}{\ell_\text{so}^{(x)}}\approx \frac{1}{\Delta_\mathrm{LH}}(2\sqrt{3}d_{v}\left(\gamma_{3}\partial_y \varepsilon_{yz}-\gamma_{2}\partial_x\varepsilon_{xz}\right) \\
    +3\gamma_{3}\alpha_Rb_{v}(\varepsilon_{xx}-\varepsilon_{yy})) \\
    \frac{1}{\ell_\text{so}^{(y)}}\approx \frac{1}{\Delta_\mathrm{LH}}(2\sqrt{3}d_{v}\left(\gamma_{3}\partial_x \varepsilon_{xz}-\gamma_{2}\partial_y\varepsilon_{yz}\right) \\
    -3\gamma_{3}\alpha_Rb_{v}(\varepsilon_{xx}-\varepsilon_{yy})),
    \label{eq:invlso}
\end{aligned}
\end{equation}
where $\alpha_R$ is the Rashba coefficient and $d_v$ is a deformation potential parameter, see Table~\ref{tab:params}.

Under a specific device geometry, we can derive compact analytical formulas for the strain pattern and hence for the spin-orbit lengths. We consider three material cases: Si/SiO$_2$, Ge/GeSi, and unstrained Ge, all with circular Al gates. We take the gate radius to be 50\,nm for Ge and 15\,nm for Si, and then extract the low-temperature strain pattern~\cite{abadillo2023hole}. The fits to analytical functions in Table~\ref{tab:compact_fits} exhibit $R^2>0.997$ against numerical strain simulations for all relevant components. 
\begin{table*}[t!]
\centering
\setlength{\tabcolsep}{3.5pt}
\renewcommand{\arraystretch}{1}
\begin{tabular}{l | l | rr | rr | rr}
\toprule
\multicolumn{2}{c|}{} & \multicolumn{2}{c|}{Si} & \multicolumn{2}{c|}{Ge} & \multicolumn{2}{c}{uGe} \\
\midrule
\multicolumn{8}{l}{\textbf{(i) Strain fitting functions}}\\
\midrule
\multicolumn{2}{l|}{\emph{Parameter units:}\quad
\(p\,(10^{-6}\,\mathrm{nm}^{-1})\), \(s\,(10^{-9}\,\mathrm{nm}^{-3})\)} &
\(p\) & \(s\) & \(p\) & \(s\) & \(p\) & \(s\) \\
$\varepsilon_{xz}$ & $p\,x + s\,x^3$
& 4.3 & 3.7 & 3.9 & -1.6 & 4.9 & -0.87 \\
$\varepsilon_{yz}$ & $p\,y + s\,y^3$
& 4.3 & 3.7 & 3.9 & -1.6 & 4.9 & -0.87 \\
\midrule
\multicolumn{2}{l|}{\emph{Parameter units:}\quad
\(p\,(10^{-7}\,\mathrm{nm}^{-2})\), \(s\,(10^{-10}\,\mathrm{nm}^{-4})\)} &
\(p\) & \(s\) & \(p\) & \(s\) & \(p\) & \(s\) \\
$\varepsilon_{xy}$ & $x\,y\,(p+s\,r^2)$
& -1.1 & -3.3 & -0.37 & 0.07 & -0.45 & -0.019 \\
$\varepsilon_{xx}-\varepsilon_{yy}$ & $(x^2-y^2)\,(p+s\,r^2)$
& 3.3 & -5.5 & -0.94 & 0.17 & -0.092 & -0.32 \\
\midrule
\multicolumn{8}{l}{\textbf{(ii) Inverse spin-orbit lengths. See Eq.~\ref{eq:invlso}}}\\
\midrule
\multicolumn{2}{l|}{\emph{Parameter units:}\quad
\(\beta_0\,(10^{-3}\,\mathrm{nm}^{-1})\), \((\beta_{ix},\beta_{iy})\,(10^{-6}\,\mathrm{nm}^{-3})\)} &
\(\beta_0\) & \((\beta_{ix},\beta_{iy})\) & \(\beta_0\) & \((\beta_{ix},\beta_{iy})\) & \(\beta_0\) & \((\beta_{ix},\beta_{iy})\) \\
$1/\ell_{\mathrm{so}}^{(x)}$ & $\beta_0 + \beta_{xx}\,x^2 + \beta_{xy}\,y^2$
& -5.3 & (3.7,\,-3.8) & 0.06 & (0.24,\,-0.31) & 2.0 & (3.1,\,-4.1) \\
$1/\ell_{\mathrm{so}}^{(y)}$ & $\beta_0 + \beta_{yx}\,x^2 + \beta_{yy}\,y^2$
& -5.3 & (-3.8,\,3.7) & 0.06 & (-0.31,\,0.24) & 2.0 & (-4.1,\,3.1) \\
\bottomrule
\end{tabular}
\caption{Functional forms of strain and spin-orbit lengths. (i) Fitting functions and parameters of the strain profile for Si, Ge, and unstrained Ge with gate radius 15, 50, and 50 nm, respectively. Every fitting function shows a correlation $R^2>0.997$. Here \(r^2=x^2+y^2\). (ii) Spin-orbit lengths functional forms extracted by applying Eq.~\eqref{eq:invlso} to the functional forms in (i). We use $\alpha_R=0$ for simplicity, and $\Delta_\mathrm{LH}=3.5$ meV, $\Delta_\mathrm{LH}=70$ meV, and $\Delta_\mathrm{LH}=3.5$ meV for Si, Ge, and unstrained Ge, respectively.}
\label{tab:compact_fits}
\end{table*}

\section{Schrieffer-Wolff transformation with Fock-Darwin states}
\label{app:fockdarwin}

The Schrieffer-Wolff (SW) transformation allows us to obtain an effective Hamiltonian projected onto a chosen subspace. Here, we extract the low-energy effective Hamiltonian of the spin-photon system restricted to the lowest orbital state. The Hamiltonian in Eq.~\eqref{eq:hybrid2D} is split as $H_\text{hyb}=\hat{H}_0+\hat{H}_I$, with
\begin{equation}
    \begin{aligned}
    \hat{H}_0&=\frac{\Pi_x^2}{2m_\parallel}+\frac{\Pi_y^2}{2m_\parallel}+V_\text{2D}(x,y)+\frac{1}{2}\mu_B\boldsymbol{\sigma}\cdot\hat{g}^{(0)}\mathbf{B}+\hbar\omega_R a^\dagger a \\
    \hat{H}_I&=\frac{1}{2}\mu_B\boldsymbol{\sigma}\cdot\delta\hat{g}\mathbf{B}+H_\text{soc}+eF_\text{zpf}(a+a^\dagger)x.
    \end{aligned}
\end{equation}
In this problem, \(\hat{H}_0\) can be solved exactly with eigenvalues \(E_n\) and eigenfunctions \(\ket{\psi_n}\), while \(\hat{H}_I\) is treated perturbatively.

We construct an anti-Hermitian operator \(\hat{S}\) and expand it order by order so that the unitary transformation~\cite{winkler2001spin}
\begin{eqnarray}\label{eq:SW transformation}
    e^{\hat{S}}He^{-\hat{S}}=\sum_{j=0}^\infty\frac{1}{j!}[H,\hat{S}]_j,
\end{eqnarray}
with \([\hat{H},\hat{S}]_{j+1}=[\hat{H},[\hat{H},\hat{S}]_j]\), \([\hat{H},\hat{S}]_0=\hat{H}\), is block-diagonal (in the eigenbasis of the unperturbed Hamiltonian \(\{\ket{\psi_n}\}\)) up to the desired order of \(\hat{H}_I\).

At first order we require \([\hat{H}_0,\hat{S}]=-\hat{H}_I\), which yields
\begin{eqnarray}\label{eq:S}
    \bra{\psi_m}\hat{S}\ket{\psi_l}=-\frac{\bra{\psi_m}\hat{H}_I\ket{\psi_l}}{E_m-E_l},
\end{eqnarray}
where \(\ket{\psi_m}\) lie in the retained subspace and \(\ket{\psi_l}\) are eliminated excited states. We restrict to first order since the perturbative energy scales are much smaller than the harmonic confinement.

To continue the derivation, it is convenient to change the orbital description from the continuous operators $x$, $y$, $\Pi_x$, and $\Pi_y$ to the Fock space. For a 2D quantum harmonic oscillator with vector potential contributions, it is possible to find a set of creation and annihilation operators $(b_+,b_-,b_+^\dagger,b_-^\dagger)$ which diagonalizes the orbital Hamiltonian ~\cite{qiong2002anisotropic}, such that we can rewrite $\hat{H}_0$ as
\begin{equation}
    \tilde{H}_0=\hbar\omega_+b_+^\dagger b_++\hbar\omega_-b_-^\dagger b_-+\frac{1}{2}\mu_B\boldsymbol{\sigma}\cdot \hat{g}^{(0)}\mathbf{B}+\hbar\omega_Ra^\dagger a,
    \label{eq:Heffh}
\end{equation}
where $\omega_\pm$ are the normal mode frequencies of the Fock-Darwin harmonic oscillator, which can be related to the original frequencies $\omega_{x,y}$, and the cyclotron frequency $\omega_B=eB_z/m_\parallel$ through the equations:
\begin{subequations}
\begin{align}
    \omega_+ &= \sqrt{\frac{\beta+\sqrt{\Delta}}{2}} \\
    \omega_- &= \sqrt{\frac{\beta-\sqrt{\Delta}}{2}} \\
   \beta &= \omega_x^2+\omega_y^2+4\omega_B^2 \\
    \Delta &= \beta^2-4\omega_x^2\omega_y^2 = \left( \omega_x^2-\omega_y^2 \right)^2+8\omega_B^2\left( \omega_x^2+\omega_y^2+2\omega_B^2 \right).
\end{align}
\label{eq:omegas}
\end{subequations}
The original operators $x$, $y$, $\Pi_x$, and $\Pi_y$ can now be related to ladder operators through a matrix operation $(x,\Pi_x,y,\Pi_y)=M\cdot(b_+,b_+^\dagger,b_-,b_-^\dagger)$, allowing us to write $\hat{H}_I$ in terms of ladder operators and heavily simplifying the perturbative calculation. The coefficients in the matrix $M$ are quite lengthy, therefore, we refer to Ref.~\cite{qiong2002anisotropic} where the calculation is performed explicitly. Using $\hat{H}_I$ in terms of ladder operators, we obtain the antihermitian operator $\hat{S}$ using Eq.~\ref{eq:S} and, by applying it to the whole Hamiltonian, we extract the effective spin-photon interaction terms from Eq.~\ref{eq:genspinphoton} up to linear order in $\mathbf{B},\omega_R$, and $\langle1/\ell_\text{so}\rangle$.

\begin{figure}
\includegraphics[width=0.7\columnwidth]{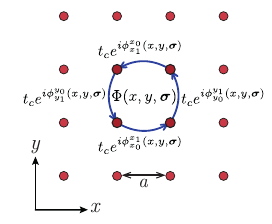}
\caption{2D Tight-binding description and inhomogeneous non-abelian phases during tunneling $\phi_{x_i}^{x_j}$ as given in Eq.~\ref{eq:peierls}, with $a$ as the lattice constant of the material, and $\Phi$ representing the flux through a plaquette.}
\label{fig:su2}
\end{figure}

\section{Tight-binding model}
\label{app:TB}
To benchmark the analytical formulas in Fig.~\ref{fig:couplingsvsL}, we consider a simple tight-binding model with lattice constant $a$ ($a^\text{Si}=\SI{5.4298}{\angstrom}$ and $a^\text{Ge}=\SI{5.6524}{\angstrom}$) and nearest-neighbor tunneling $t_{ij}$ based on the discretization of Eq.~\ref{eq:2dbase}. We include the spin-orbit interaction $H_\text{soc}$ and the magnetic vector potential via the Peierls substitution
\begin{equation}
    t_{ij}\rightarrow t_{ij}\exp\left[i\int_{\mathbf{x}_i}^{\mathbf{x}_j}\left(\frac{q}{\hbar} \mathbf{A}(x,y)\sigma_0+\frac{1}{2}\boldsymbol{\sigma}\cdot\hat{\mathcal{A}}(x,y)\right)\cdot d\mathbf{l}\right],
    \label{eq:peierls}
\end{equation}
where $\mathbf{A}$ is the electromagnetic vector potential and $\hat{\mathcal{A}}(x,y)$ is a spin-orbit matrix that couples spin and real space, with components $\hat{\mathcal{A}}_{\mu}^{\,\nu}(x,y)=\frac{1}{\ell_\text{so}^{(\nu)}(x,y)}\,n^{(\nu)}_{\!so,\mu}(x,y)$. We generally use the symmetric gauge but have verified other gauges yield identical results, as expected by gauge invariance. In Fig.~\ref{fig:su2}, we show a schematic of this tight-binding simulation including the Peierls Abelian and non-Abelian phases.

For a given working point (magnetic field and harmonic lengths), we compute the eigenvalues and spin eigenvectors $\ket{0}$ and $\ket{1}$. We include a classical drive term $eF_\text{zpf}x$, which is then rotated to the eigenbasis and projected onto the spin subspace, yielding a dipole operator of the form $\hat{d}=2\,\boldsymbol{\lambda}\cdot\boldsymbol{\sigma}$, where $\boldsymbol{\lambda}$ coincides with the spin-photon coupling vector.

\bibliography{bib}

\end{document}